\documentclass[1p]{elsarticle}

\usepackage{graphicx,epsfig}
\usepackage{amsmath,mathrsfs,amsfonts}
\usepackage {amssymb}
\usepackage {longtable}
\usepackage{multirow}
 \usepackage{enumerate}

\usepackage{bm}
\usepackage{amsfonts}
\usepackage{subfigure}
\usepackage{color}
\usepackage{relsize}
 \usepackage[utf8]{inputenc}
\usepackage{hyperref}

\newcommand{\be}{\begin{equation}}
\newcommand{\ee}{\end{equation}}
\newcommand{\bea}{\begin{eqnarray}}
\newcommand{\eea}{\end{eqnarray}}

\journal{Journal of \LaTeX\ Templates}

\biboptions{sort&compress}

\begin{document}

\begin{frontmatter}

\title{Inflation in a  scalar-vector gravity theory}


\author{Manuel Gonzalez-Espinoza\fnref{myfootnote}}

\address{Laboratorio de investigaci\'on de C\'omputo de F\'{i}sica, Facultad de Ciencias Naturales y Exactas,
Universidad de Playa Ancha, Subida Leopoldo Carvallo 270, Valpara\'{i}so, Chile}
\author{Ram\'on Herrera\fnref{myfootnote2}}
\address{Instituto de F\'{\i}sica, Pontificia Universidad Cat\'olica de 
Valpara\'{\i}so, 
Casilla 4950, Valpara\'{\i}so, Chile}
\fntext[myfootnote]{manuel.gonzalez@pucv.cl}
\fntext[myfootnote2]{ramon.herrera@pucv.cl}

\begin{abstract}
We study the possibility that inflation is driven by a scalar field together with a vector field minimally coupled to gravity. By assuming an effective potential that incorporates both fields into the action, we explore two distinct scenarios: one where the fields interact and another where they do not. In this context, we find different analytical solutions to the background scalar-vector fields dynamics during the inflationary scenario considering the slow-roll approximation. Besides, general conditions required for these models of two fields to be realizable are determined and discussed. From the cosmological perturbations, we consider a local field rotation, and then we determine these perturbations (scalar and tensor) during inflation, and we also utilize recent cosmological observations for constraining the parameter-space in these scalar-vector inflationary models.

\end{abstract}

\end{frontmatter}

\section{Introduction}\label{Introduction}

Inflation has emerged as a fundamental paradigm for describing the early universe’s physics. However, the identity of the field(s) responsible for driving inflation remains an open question. Future cosmological observations, particularly those related to the Cosmic Microwave Background (CMB), are hoped to help narrow down the possibilities and provide insights into the underlying physics.

The incorporation of scalar fields into the inflation framework is motivated by the quest for a dynamic driving force behind the accelerated expansion. The simplest inflationary scenario involves a canonical scalar field, denoted as $\phi$, which acts as the inflaton \cite{Guth:1980zm, Linde:1981mu, Mukhanov:2005sc}. These scalar fields introduce a versatile energy component that can evolve over time. Its dynamical nature allows for variations in the equation of state during different epochs of inflation, offering a diverse landscape for exploring inflationary scenarios \cite{Caldwell:1997ii, Vernizzi:2004nc, Langlois:1999dw, Langlois:2008mn, Seery:2005gb, ArkaniHamed:2003uz}. This adaptability is crucial for constructing models that can be fine-tuned to match the increasingly precise observational data we gather from various cosmological surveys.

Notably, inflation based on scalar field offers a simple explanation for the origin of Cosmic Microwave Background (CMB) temperature anisotropies. Simultaneously, it provides a mechanism to account for the Large-Scale Structure (LSS) of our Universe by generating primordial density perturbations via quantum fluctuations during the inflationary epoch \cite{Baumann:2009ds, Lyth:2009zz, Mukhanov:1990me, Linde:1982zj}. These fluctuations are stretched to macroscopic scales by the rapid expansion, seeding the formation of galaxies and clusters observed today.

In the context of single-field inflation, curvature perturbation remains conserved on large scales. However, in multi-field inflation scenarios, the curvature perturbation can undergo significant modifications on large scales due to its sourcing from entropy (or isocurvature) perturbations. This intriguing feature was initially highlighted within the framework of Jordan-Brans-Dicke gravity, where the gravitational sector incorporates a scalar field \cite{Starobinsky:1994mh}. This effect has further been illustrated within specific inflationary models based on string theory constructions \cite{Lalak:2007vi, Brandenberger:2007ca}. Consequently, adhering strictly to an effectively single-field scenario, despite its appealing simplicity, may sometimes lead to misleading conclusions. The final curvature perturbation, which will eventually be observed, can occasionally originate predominantly from the entropy modes. In the context of multi-field inflation, the intriguing possibility—depending on the reheating scenario—exists to generate both adiabatic and isocurvature perturbations after the inflationary phase. This correlation was pointed out in the literature \cite{Langlois:1999dw, Langlois:2008mn, Seery:2005gb}.

{
Recent studies in modified gravity frameworks have demonstrated the potential of scalar fields coupled with higher-order curvature corrections to produce complex inflationary dynamics that align well with current observational data \cite{EurPhysJC2023, Symmetry2023}. These models extend beyond the limitations of the single-field paradigm, enabling more flexibility in constructing inflationary scenarios that can be fine-tuned to match the precise measurements of the CMB \cite{EurPhysJC2023}.
The incorporation of quantum effects and loop corrections in scalar field-driven inflationary models, as discussed in \cite{AnnPhys2020, JHEP2018}, provides further stability to the inflationary attractors and bridges the gap between classical General Relativity and quantum field theoretical perspectives. This approach allows for the exploration of inflationary solutions that maintain consistency with the early-universe conditions and align with effective field theories at high energies \cite{JHEP2018}.
Scalar fields in modified gravity frameworks can lead to unique observational signatures that distinguish them from standard inflationary models. These signatures include specific features in the CMB polarization and potential gravitational wave imprints, which offer new avenues for testing the validity of these models through upcoming observational campaigns \cite{Symmetry2023, PhysRevD104}. These models enrich the theoretical landscape by incorporating scalar-tensor interactions that can explain the generation of primordial perturbations and anisotropies, as well as the evolution of large-scale structures.
As the field progresses, the study of scalar field inflation within modified gravity remains a promising route for addressing unresolved questions about the mechanisms driving the universe's accelerated expansion and the formation of its structure. Enhanced by insights from both observational and theoretical advancements, these frameworks provide a deeper understanding of the potential dynamics at play during the inflationary epoch and contribute to a more comprehensive picture of early-universe cosmology \cite{EurPhysJC2023, AnnPhys2020, JHEP2018}.
}

On the other hand, the motivation to introduce vector fields within the cosmic inflation framework is rooted in the aspiration to broaden our theoretical perspective beyond scalar fields \cite{Golovnev:2008cf}. The inclusion of vector fields in inflationary models can lead to distinctive observational predictions that differ from those of scalar-driven inflation. These predictions may offer valuable insights into the underlying physics of the inflationary epoch and provide a means of discriminating between different inflationary scenarios \cite{dimopoulos2009statistical}.

{The motivation behind studying inflation driven by a combination of a scalar field and a cosmic triad—a set of three orthogonal vector fields minimally coupled to gravity—stems from the need to explore extensions of the standard single-field inflationary paradigm. While single-field models have successfully explained the large-scale homogeneity and isotropy of the universe \cite{Guth:1980zm,Linde:1981mu}, they struggle to address potential small deviations from isotropy and generate specific types of non-Gaussianities \cite{Planck:2018jri}. The inclusion of a cosmic triad, which naturally preserves rotational symmetry and isotropy on large scales while allowing for directional dynamics locally \cite{Armendariz-Picon:2004say,Golovnev:2008cf}, introduces a new degree of freedom that can enrich the inflationary framework. The cosmic triad configuration is particularly intriguing due to its stability properties and its potential to generate anisotropic features without breaking the overall isotropy of the inflationary expansion \cite{Dimopoulos:2008yv}. By coupling this setup with a scalar field, the model can leverage the benefits of traditional scalar-driven inflation while simultaneously exploring vector field dynamics. This opens up new possibilities for understanding how interactions between different types of fields could have influenced the universe's evolution during the inflationary epoch \cite{Maleknejad:2011sq,Bartolo:2013msa}.}

Vector fields offer solutions to some of the shortcomings or unanswered questions in traditional scalar-driven inflationary models. They have the potential to address issues such as the generation of primordial magnetic fields, the origin of large-scale structures, and the generation of anisotropies in the cosmic microwave background radiation \cite{dimopoulos2009statistical}. By exploring the dynamics and implications of vector fields in cosmic inflation, we can advance our understanding of the early universe. Investigating the role of vector fields in driving inflationary expansion allows us to gain deeper insights into the fundamental processes that shaped the cosmos during its formative stages \cite{golovnev2009vector}.

The coupling between scalar and vector fields introduces distinctive observational consequences that serve as probes for validating or constraining inflationary models. Constraints derived from cosmic microwave background (CMB) observations, such as those obtained by the Planck satellite mission \cite{Planck2018, Ade:2015lrj}, play a crucial role in delineating the parameter space of scalar-vector coupling scenarios. Additionally, analyses of large-scale structure surveys, such as the Sloan Digital Sky Survey (SDSS) \cite{Alam:2016hwk, Ross:2014qpa}, offer complementary insights into the impact of scalar-vector interactions on the clustering properties of galaxies and dark matter halos.

Moreover, gravitational wave experiments, such as those conducted by the Laser Interferometer Gravitational-Wave Observatory (LIGO) \cite{Abbott:2016blz, Lerner:2003ey}, provide unique opportunities to probe the cosmic fabric for signatures of primordial gravitational waves generated during inflation. Scalar-vector coupling dynamics can leave distinct imprints on the stochastic gravitational wave background, offering a tantalizing prospect for direct observational validation of inflationary scenarios featuring coupled fields.

Extensions of the standard inflationary paradigm, such as multi-field inflation models \cite{Langlois:1999dw, Langlois:2008mn, Seery:2005gb}, motivate to incorporate scalar-vector interactions to explore the rich phenomenology of early-universe dynamics. By incorporating both scalar and vector degrees of freedom, these frameworks offer novel avenues for addressing fundamental questions regarding the origin of cosmic structure and the nature of dark energy.

Furthermore, cosmological simulations utilizing state-of-the-art computational techniques \cite{Tye:2005fn} enable detailed investigations into the formation and evolution of cosmic structures within scalar-vector coupled inflationary scenarios. These simulations provide invaluable insights into the nonlinear dynamics of cosmic strings, their gravitational effects, and their observational implications, fostering a deeper understanding of the cosmic ballet orchestrated by scalar and vector fields during the inflationary epoch.

In summary, the exploration of scalar-vector coupling within the inflationary paradigm represents a frontier in theoretical cosmology, offering a fertile ground for interdisciplinary research at the interface of particle physics, astrophysics, and cosmology. As observational data improves in quality and quantity, ongoing efforts to refine and confront scalar-vector coupled inflationary models with empirical evidence will undoubtedly enrich our understanding of the early universe and its intricate tapestry of fundamental interactions.\\

The outline of the paper is as follows: The next section shows the background dynamics considering the slow roll approximation in a scalar-vector gravity theory.  In section 3, we analyze the cosmological perturbations for double inflation, and we obtain explicit expressions for the scalar power spectrum, spectral index, and tensor-to-scalar ratio under a local field rotation. In section 4, we study the first example of our model, in which we consider a specific potential in which there is no interaction between the fields. In section 5, we analyze a second example, in which we study an effective potential that presents an interaction between the scalar field and the vector field. Finally, section 6 resumes our results and exhibits
our conclusions. We chose units so that $c =\hbar= 1$.

\section{ Scalar Vector gravity theory}\label{Multi_Infl_}

{We start with the action for the Maxwell scalar  gravity theory given by \cite{Armendariz-Picon:2004say}. This choice is particularly significant because the cosmic triad model benefits from the symmetrical properties imparted by the Maxwell term. These properties facilitate the isotropic contributions of the vector fields to the expansion on large scales while preserving the potential for anisotropic perturbations at smaller scales \cite{Armendariz-Picon:2004say,Maleknejad:2011sq}. This combination is valuable for generating distinctive signatures in the cosmic microwave background (CMB), such as specific types of non-Gaussianity and tensor modes, which can be constrained by observational data \cite{Planck:2018jri}. Moreover, employing a Maxwell-like structure for the vector fields aligns with the broader aim of constructing a model where the scalar and vector fields interact seamlessly. The triad configuration naturally stabilizes the inflationary dynamics and allows for complex field interactions that can extend the predictability and explanatory power of traditional inflation models \cite{Dimopoulos:2008yv}. In summary, the Maxwell scalar gravity theory was chosen for its simplicity and stability and its critical role in supporting the cosmic triad configuration. This ensures that the action can model a set of orthogonal vector fields contributing to the inflationary process in a theoretically consistent and observationally testable manner. In this context, the action for the Maxwell scalar gravity theory is given by  \cite{Armendariz-Picon:2004say}
\bea
&S=&\int \text{d}^{4}{x} \sqrt{-g}\Bigg[\dfrac{R}{2\kappa^2} + X_1 -\sum_{a=1}^3 \dfrac{1}{4}{F^a}_{\mu\nu}{F^a}^{\mu\nu}- V(\phi,X_2)\Bigg], \label{action00}
\eea 
}
where $\kappa^2=8\pi G$, with $G$ being the gravitational constant and $\phi$ denotes 
the scalar field  along with its kinetic term $X_1$. The quantity ${F^a}_{\mu\nu}$ corresponds to the antisymmetric field strength tensor and it is defined as ${F^a}_{\mu\nu}=\partial_\mu A^a_\nu-\partial_\nu A^a_\mu$ where  $A^a_\mu$ is a four-vector potential with the mass term $X_2$. It is important to note that the superscript $a$ represents each vector field that makes up the cosmic triad. Thus, we have 
\begin{equation}
    X_1 = - \dfrac{1}{2} \nabla_\mu \phi \nabla^\mu\phi, \ \ \ \    X_2 = \dfrac{1}{6}\sum_{a=1}^3 g^{\mu\nu}A^a_\mu A^a_\nu = \dfrac{1}{6}\sum_{a=1}^3{A^a\,^2} ,
\end{equation}
Additionally, the quantity $V(\phi,X_2)$ denotes the  effective potential associated to the scalar field $\phi$ and the  field $X_2$, respectively.

{While the primary focus of this paper is on analyzing the field equations and inflationary dynamics resulting from this action, it is worth noting that the structure of action \eqref{action00} suggests the potential for deriving a canonical Hamiltonian \cite{Armendariz-Picon:2004say,Golovnev:2008cf,Maleknejad:2011sq}. This would facilitate a more detailed phase space analysis and stability studies. Although the explicit Hamiltonian derivation lies outside the scope of this work, such an approach would provide an additional layer of understanding and could be explored in future research.\\}

{By assuming that the vector field as $A^a_\mu=\delta^a_\mu A(t)a(t)$
together with  a homogeneous scalar field $\phi=\phi(t)$ and considering a 
spatially flat Friedmann Robertson Walker (FRW) metric,  we obtain from the action given by Eq.(\ref{action00}) the following  background equations }

\begin{eqnarray}
\frac{3 H^2}{\kappa ^2}&=&  \frac{3}{2} (A H + \dot{A})^2  + \dfrac{1}{2}\dot{\phi}^2 + V ,\label{beq1}\\
-\frac{2 \dot{H}}{\kappa ^2}&=& - A^2 V_{,X_2} + 2 (A H + \dot{A})^2+ \dot{\phi}^2,\label{beq2}
\end{eqnarray}
 \begin{equation}
\ddot{\phi}+ 3 H \dot{\phi } + V_{, \phi} =0, \label{beq3}
 \end{equation}
 and 
 \begin{equation}
   \dfrac{d (A H + \dot{A})}{d t}+ 2 H (A H + \dot{A}) = A V_{,X_2}, \label{beq4}
\end{equation}
here, we have considered that $X_2=A^2/2$. Besides, the quantity $H=\dot{a}/a$ corresponds to the Hubble parameter, where $a$ denotes the scale factor.
{Additionally, we have considered a flat FRW universe, which simplifies the study of cosmic dynamics during the inflationary scenario and the analysis of cosmological perturbations. }

In the following, we will utilize that the dots correspond differentiation with respect to the time, and  we will use that the notation $V(\phi,X_2)=V$, $V_{,\phi}$ represents $V_{,\phi}=\partial V/\partial \phi$, 
$V_{,X_2}$ to $V_{,X_2}=\partial V/\partial X_2$, etc.

{ In the following, we will work under the slow-roll approximation for both scalar fields $\phi$ and $X_2$. However, this is not a strict requirement; one of the fields could, in principle, roll quickly to the minimum of its potential, simplifying the problem to single-field inflation. However, we aim to consider the more general case where both fields undergo slow-roll, following the analysis developed in Refs.\cite{Mollerach:1991qx,Garcia-Bellido:1993fsr,Barrow:1995fj,Vernizzi:2006ve}.
In this context, 
in order to study the inflationary scenario, we can  introduce the following 
slow-roll parameters:}
\begin{equation}
    \epsilon_\phi = \dfrac{1}{2 \kappa^2} \left( \dfrac{V_{,\phi}}{V} \right)^2 , \ \ \ \ \ \ \ \ \eta_{\phi \phi} = \dfrac{1}{\kappa^2} \left( \dfrac{V_{,\phi \phi}}{V} \right) , \label{pa1}
\end{equation}
\begin{equation}
    \epsilon_A = \dfrac{1}{2 \kappa^2} \left( \dfrac{A V_{,X_2}}{V} \right)^2 , \ \ \ \ \ \ \ \ \eta_{A A} = \dfrac{1}{\kappa^2} \left( \dfrac{A^2 V_{,X_2 X_2}}{V} \right) , \,\,\,\,\,\mbox{and}\ \ \ \  \eta_{\phi A} = \dfrac{1}{\kappa^2} \left( \dfrac{A V_{,\phi X_2}}{V} \right)  ,\label{pa2}
\end{equation}
respectively. 
{Here, from Eq.(\ref{pa1}) we note that  the first two conditions are simply the expected generalization of the slow-roll conditions for a single field  and in this case the field $\phi$. Similarly, from Eq. (\ref{pa2}), the first two parameters associated with the scalar 
$A$ arise by ensuring that both the conditions on the effective potential and the field 
$A$ also exhibit slow-roll behavior. The last parameter of Eq.(\ref{pa2})  $\eta_{\phi A}$ ensures the slow roll for both fields. Thus, the slow-roll approximation related to inflation corresponds to a fairly flat potential
and the slow-roll approximation invokes a hierarchy of
dimensionless ratios in terms of the derivatives associated to both fields.}

{
In relation to the inflationary stage under the  slow roll approximation, we can comment that there is a crucial difference between single field  and inflation
with two or many scalar fields\cite{Contreras:1995uv,Garcia-Bellido:1995him}. It is well known  that  in the case in which we have a single field  the slow-roll solution forms a
one-dimensional phase space. This means that once the inflationary attractor has been reached, there is only one unique trajectory. In particular, the end
of inflation takes place at a fixed value of the inflaton field which in turn
corresponds to a fixed energy density. However, if two fields are present,
the phase-space becomes two-dimensional and then there is an infinite number of
possible classical trajectories in field space. The values of the two fields at
the end of inflation will in general depend on the choice of trajectory.}

Thus, under the slow-roll approximation, these parameters are much less than the unit, and then 
the background equations are reduced to
\begin{eqnarray}
\frac{3 H^2}{\kappa ^2}&\simeq&   V, \label{srbeq1}\\
 3 H \dot{\phi }&\simeq&  -V_{, \phi} , \label{srbeq2} \\
 2 H (A H + \dot{A})&\simeq&  A V_{,X_2}.   \label{srbeq3}
\end{eqnarray}
Also, we note that the Eqs.(\ref{srbeq2}) and (\ref{srbeq3}) can be rewritten in terms of the slow roll parameters as

\begin{equation}
    \dot{\phi}^2 \simeq	\dfrac{2}{3} \epsilon_\phi V ,\,\,\,\,\,\,\mbox{and}\ \ \ \ \ \ \ \ (A H + \dot{A})^2 \simeq	\dfrac{3}{2} \epsilon_A V ,
\end{equation}
respectively.

Additionally, from equation \eqref{beq2} and considering that the parameter $\epsilon=-\dot{H}/H^2$, we can write

\begin{equation}
    \epsilon \simeq -\dfrac{3}{\sqrt{2}} A \kappa \sqrt{\epsilon_A} + \dfrac{9}{2} \epsilon_A + \epsilon_{\phi}. \label{slow_roll_epsilon}
\end{equation} 

By combining the equations under the slow roll approximation (\ref{srbeq1})-(\ref{srbeq3}), we can find a relation between the scalar field $\phi$ and the scalar $A$ associated to the vector field  given by
\begin{equation}
    \dfrac{d \phi}{d \ln A} \simeq - \dfrac{2}{3 } \left(\dfrac{V_{,\phi}}{V_{,X_2}-\frac{2}{3}\kappa^2 V}\right) .\label{FA}
\end{equation}
In this way, depending on the chosen effective potential $V= V(\phi,X_2)=V(\phi,A^2/2)$, we could obtain an analytical solution to Eq.(\ref{FA}) and then we could find the relation $\phi=\phi(A)$ or vice versa, i.e., $A=A(\phi)$.

\section{Cosmological Perturbations}
In this section, we will analyze the cosmological perturbations in our model. Following Refs.\cite{Taruya:1997iv,Wands:2002bn}, we have that the perturbed Klein-Gordon equations by considering a perturbed  FRW metric spacetime become

$$
\delta\ddot{ \phi} + 3 H \delta\dot{ \phi} + \left[\dfrac{k^2}{a^2}+V_{\phi \phi} - \dfrac{\kappa^2}{a^3} \dfrac{d}{d t} \left( \dfrac{a^3}{H} \dot{\phi}^2\right) \right]  \delta\phi 
$$
\begin{equation}
= \left[ A V_{,\phi X_2}+3 \dfrac{\kappa^2}{a^3} \dfrac{d}{d t} \left( \dfrac{a^3}{H} \dot{\phi}(A H + \dot{A})\right)\right]\delta A,  
\end{equation}
and
$$
3\delta\ddot{ A} + 9 H \delta\dot{ A} + \left[ \dfrac{k^2}{a^2}+A^2 V_{,X_2 X_2} - 3\dfrac{\kappa^2}{a^3} \dfrac{d}{d t} \left( \dfrac{a^3}{H} (A H + \dot{A})^2\right) \right]  \delta A 
$$
\begin{equation}
= \left[ A V_{,\phi X_2} +3 \dfrac{\kappa^2}{a^3} \dfrac{d}{d t} \left( \dfrac{a^3}{H} \dot{\phi}(A H + \dot{A})\right)\right]  \delta \phi,  
\end{equation}
respectively.

By considering the linear perturbation on large scales in which $k\gg aH$ and neglecting those terms which include second order time derivatives and using the slow roll parameters given by Eqs.(\ref{pa1}) and (\ref{pa2}), we find that the perturbed Klein-Gordon equations can be approximate to
\begin{equation}
 H^{-1} \delta\dot{ \phi} + \left[\eta_{\phi \phi} - 2 \epsilon_{\phi} \right]  \delta\phi \simeq\left[ \eta_{\phi A} \pm 9 \sqrt{\epsilon_\phi \epsilon_A}\right]  \delta A , 
 \end{equation}
 and
 \begin{equation}
3 H^{-1}\delta \dot{ A} + \left[ \eta_{AA} - \dfrac{27}{2} \epsilon_A \right]  \delta A \simeq \left[ \eta_{\phi A} \pm 9 \sqrt{\epsilon_\phi \epsilon_A} \right]  \delta \phi.  
\end{equation}
 
 {It is well known that in the simplest models of inflation driven by a single scalar field, fluctuations associated with this field generate a primordial adiabatic spectrum. The amplitude of this spectrum can be described by the comoving curvature perturbation, which remains constant on super-Hubble scales.  However, when more than one scalar field is present, the role of non-adiabatic fluctuations must also be considered. On the one hand, these fluctuations can influence    the evolution of the curvature perturbation (adiabatic perturbation) and may also lead to the generation of isocurvature (or called entropy) perturbations after inflation. Thus, following Refs.\cite{Wands:2002bn,Gordon:2000hv}, in multi-field inflation models, the evolution of both curvature and isocurvature perturbations can be analyzed by decomposing field perturbations into components along the background trajectory in field space 
  or adiabatic field perturbation and components orthogonal to the background trajectory (entropy field perturbation). In this context, }
 for the case of the scalar perturbations in a system with two fields (double inflation) and following Refs.\cite{Wands:2002bn,Gordon:2000hv}, we must utilize a local field rotation in order to recognize the adiabatic and entropy perturbations in the orthogonal form to the background trajectory in field space. Thus, considering this rotation on the perturbed quantities, we can define two quantities $\delta \sigma$ and $\delta s$, in which $\sigma$ represents the ``adiabatic field" and $s$ denotes the ``entropy field" such that\cite{Gordon:2000hv}
\begin{eqnarray}
 \delta \sigma =\cos{\theta} \delta \phi + \sin{\theta} \delta A , \,\,\,\,\,\,\,\,\,\mbox{and}
 \,\,\,\,\,\,\,\,\delta s = - \sin{\theta} \delta \phi + \cos{\theta} \delta A,
\end{eqnarray}
where the angle $\theta$ is related to the slow roll parameters from the relation $\tan \theta =(A H + \dot{A})/\dot{\phi} =a^{-1} (aA\dot{)}/\dot{\phi}=( 3/2) \sqrt{\epsilon_A/\epsilon_\phi}$. Under this rotation, the curvature and entropy perturbations during the inflationary epoch can be written in standard form as; 
$
    \mathcal{R} \simeq \dfrac{H \delta\sigma}{\dot{\sigma}} \mbox{and}\,\,\mathcal{S} \simeq \dfrac{H \delta s}{\dot{\sigma}}
$, respectively\cite{Wands:2002bn,Gordon:2000hv}.

From this rotation, we have that the slow-roll parameter associated to the slope orthogonal to the trajectory, $\epsilon_s \simeq 0$, and the new slow-roll parameters related to $\sigma$ and $s$ fields can be written as \cite{Wands:2002bn}  
\begin{eqnarray}
 \eta_{\sigma s} &=& \frac{1}{3} \eta _{\text{AA}} \sin ^2(\theta )-\frac{4}{3} \eta _{\text{$\phi $A}} \sin (\theta ) \cos (\theta )+\eta _{\phi \phi } \cos ^2(\theta ) ,\\
 \eta_{\sigma\sigma} &=& \frac{1}{3} \eta _{\text{AA}} \sin (\theta ) \cos (\theta )+\frac{1}{3} \eta _{\text{$\phi $A}} \sin ^2(\theta )-\eta _{\text{$\phi $A}} \cos ^2(\theta )-\eta _{\phi \phi } \sin (\theta ) \cos (\theta ), \\
 \eta_{s s} &=& \frac{1}{3} \eta _{\text{AA}} \cos ^2(\theta )+\frac{4}{3} \eta _{\text{$\phi $A}} \sin (\theta ) \cos (\theta )+\eta _{\phi \phi } \sin ^2(\theta ) , 
\end{eqnarray}
and the parameter associated to the slope of $V_{,\sigma}$ is defined as
\begin{equation}
    \epsilon = \dfrac{1}{2 \kappa^2} \left( \dfrac{V_{,\sigma}}{V} \right)^2 \simeq -\dfrac{3}{\sqrt{2}} A \kappa \sqrt{\epsilon_A} + \dfrac{9}{2} \epsilon_A + \epsilon_{\phi}.
\label{ee}
\end{equation}

Besides, from this rotation, the  background equations can be rewritten as
\begin{equation}
    \dot{\sigma}^2 \simeq \dfrac{2}{3} \epsilon\, V , \ \ \ \ \ \ \ H^{-1} \dot{\theta} \simeq - \eta_{\sigma s},
\end{equation}
and the adiabatic and entropy perturbations satisfy the following equations:
\begin{eqnarray}
 H^{-1} \delta\dot{ \sigma} &\simeq& (2 \epsilon - \eta_{\sigma \sigma})\delta\sigma - 2 \eta_{\sigma s} \delta s , \\
 H^{-1} \delta\dot{ s} &\simeq& - \eta_{ss} \delta s.
\end{eqnarray}

In this form, following Refs.\cite{Wands:2002bn,Gordon:2000hv},  the adiabatic and entropy power spectrum at Hubble-crossing when $k=a_*H_*$ can be defined   as 
\begin{equation}
     \mathcal{P}_{\mathcal{R}} |_* \simeq \mathcal{P}_{\mathcal{S}} |_* \simeq \left(\frac{H^2}{2\pi\dot{\sigma}}\right)_*^2\simeq\dfrac{8}{3 \epsilon} \kappa^4 V_*.\label{PR}
\end{equation}
In what follows, the subscript $*$ is utilized  to denote the epoch in which the cosmological
scale exits the horizon.

Also, the scalar spectral index $n_{\mathcal{R}}$ associated to the adiabatic perturbation together with the scalar spectral index $n_{\mathcal{S}}$ related to the entropy spectrum at the Hubble crossing become \cite{Wands:2002bn}
\begin{equation}
    n_{\mathcal{R}} |_* \simeq n_{\mathcal{S}} |_* \simeq - 6 \epsilon_* + 2 \eta_{{\sigma \sigma}_{*}}.
\label{nR}
\end{equation}
Here the scalar spectral index $n_{\mathcal{R}}$ and $n_{\mathcal{S}}$ are defined as $n_{\mathcal{R}}=d \ln \mathcal{P}_{\mathcal{R}}/d \ln k $ and $n_{\mathcal{S}}=d \ln \mathcal{P}_{\mathcal{S}}/d \ln k $, respectively.

{
In relation to the generation of tensor perturbation $\mathcal{P}_{T}$ during the inflationary scenario, it remains unaltered  in the framework of double inflation \cite{Liddle:2000cg,Wands:2002bn}. 
The analysis of the evolution of tensor perturbations or primordial gravity waves can also be reduced
to the study of a parametric oscillator. In this sense, the amplitude of each transverse Fourier mode of the gravity wave denoted by $\mu_{\bf{k}}(\tau)$ (with $\tau$ the conformal time) satisfies the equation $\mu_{\bf{k}}''+(k^2-a''/a)\mu_{\bf{k}}=0$. In this form, noting that the power spectrum associated with the gravitational wave, defined by the two-point correlation function, is  $\mathcal{P}_{T}(k)=(2k^3/\pi^2)\lvert \mu_{\bf{k}}/a \rvert^2 $
and considering   the slow-roll approximation  we can write \cite{Liddle:2000cg,Riotto:2002yw,Wands:2002bn}
\begin{equation}
    \mathcal{P}_{T}|_* \simeq 128\kappa^2\,H_*^2\simeq \dfrac{128}{3} \kappa^4 V_*=\dfrac{128}{3M_{Pl}^4}  V_*\,,
\end{equation}
where we have considered that $\kappa^2=8\pi G=M_{Pl}^{-2}$ with $M_{Pl}$ the Planck mass \cite{Wands:2002bn}.}

The tensor spectral index $n_T$ associated to the tensor perturbation is expressed   in terms of
the slow-roll parameter $\epsilon$
as $ n_T|_*=d\ln \mathcal{P}_{T}/d\ln k|_*=-2\,\epsilon$. 

Additionally, 
an important observational quantity corresponds to the tensor-to-scalar ratio, denoted as  $r$.  This observational parameter  at Hubble-crossing can be written as
\begin{equation}
   r |_*= \left( \dfrac{\mathcal{P}_T}{\mathcal{P}_{\mathcal{R}}} \right)_* \simeq 16 \epsilon|_* \simeq - 8 n_T|_*.\label{rr}
\end{equation}

In the following, we will explore some inflationary stages within the framework of a scalar vector gravity theory. Specifically, we will investigate the inflationary scenario, assuming two types of effective potentials, $V(\phi,X_2)$, associated with double inflation. Firstly, we will analyze a separable potential of the form $V(\phi,X_2)= \bar{V}(\phi)+\tilde{V}(X_2)$, in order to describe an inflationary stage, where there is no interaction between the fields $\phi$ and $X_2$. In a second scenario, we will analyze a potential $V$, in which the fields interact by mean of  the product  of the potentials $\bar{V}(\phi)$ and $\tilde{V}(X_2)$,  
such that  the effective potential $V(\phi,X_2)$ is defined as $V(\phi,X_2)= \bar{V}(\phi)\times\tilde{V}(X_2)$.


\section{Example I: Effective Potential $V(\phi,X_2) =  \bar{V}(\phi)+\tilde{V}(X_2) $}

In this section, we will assume a specific potential in which there is no interaction between the fields. The effective potential (separable potential) can be expressed as the sum of $\bar{V}(\phi)$ and $\tilde{V}(X_2)$ such that \cite{Vernizzi:2006ve,Battefeld:2006sz}
\begin{equation}
V(\phi,X_2)= \bar{V}(\phi)+\tilde{V}(X_2),
\end{equation}
where $\bar{V}(\phi)$ and $\tilde{V}(X_2)$ are two arbitrary potentials that exclusively depend on the scalar field $\phi$ and $X_2$, respectively. From Eq.(\ref{FA}), we find that the relation between both fields can be determined from the equation
\begin{equation}
    \dfrac{d \phi}{d \ln X_2} = - \dfrac{1}{3 }\left( \dfrac{\bar{V}_{,\phi}}{\tilde{V}_{,X_2}-\frac{2}{3}\kappa^2 \tilde{V} -\frac{2}{3}\kappa^2 \bar{V}}\right) , \label{relation_sum}
\end{equation}
and the number of e-folds $N$ defined as $dN=Hdt$ results
\begin{equation}
    d N = - \kappa^2 \dfrac{\bar{V}}{\bar{V}_{,\phi}} d \phi + \dfrac{\kappa^2}{3 X_2} \dfrac{\tilde{V}}{(\tilde{V}_{,X_2}-\frac{2}{3}\kappa^2 \tilde{V} -\frac{2}{3}\kappa^2 \bar{V}) } d X_2 . 
\end{equation}
In order to derive an analytical expressions for this separable potential, we can consider that the effective potentials  for  $\bar{V}(\phi)$ and $\tilde{V}(X_2)$ are exponential potentials defined as
\begin{equation}
\bar{V}(\phi)= \bar{V}_0
e^{\alpha_1 \phi^2},\,\,\,\,\,\,\mbox{and}\,\,\,\,\,\,\tilde{V}(X_2) = \tilde{V}_0 e^{\alpha_2\,X_2},\label{VS}
\end{equation}
respectively. Here $\bar{V}_0$ and $\tilde{V}_0$ are two arbitrary constants, representing the amplitude of their respective potentials, and $\alpha_1$ and $\alpha_2$ correspond to two constants with dimensions of $\kappa^2$ (or $M_{Pl}^{-2}$, where $M_{Pl}$ is the Planck mass). The exponential potentials in the context of the early universe have been extensively investigated in the literature, see e.g.,  Refs.\cite{Copeland:1997et,Lucchin:1984yf}

In order to find an analytical relation between the fields $\phi$ and $X_2$, we consider the special case in which the constant $\alpha_1=(3/4)\alpha_2=\kappa^2/2$. 
Thus, combining Eqs.(\ref{relation_sum}) and (\ref{VS}), we find that the relation between the fields $\phi$ and $X_2$ is given by
\begin{equation}
   \dfrac{d \phi}{d X_2} = \dfrac{\phi}{2 X_2}\,\,\,\,\,\,\,\Rightarrow\,\,\,\,\,\,\,\,\,\,\,
 X_2 = X_2^{(0)} \phi^2 , \label{relation_sum_f}
\end{equation}
where $X_2^{(0)}$ corresponds to a new integration constant (dimensionless).

Introducing the number of e-folds $N(t=t_*)=N_*$ between two values of times $t_*$ and $t_f$  (or between two values of the scalar fields) we have
\begin{equation}
     N_*=   \int_{t_*}^{t_f} Hdt \simeq \int_{\phi_f}^{\phi_*} \dfrac{1}{\phi} d \phi - \dfrac{1}{2} \dfrac{\tilde{V}_0}{\bar{V}_0} \int_{X_{2f}}^{X_{2*}} \dfrac{1}{X_2} \dfrac{e^{\frac{2}{3} \kappa^2 X_2}}{e^{\frac{\kappa^2 \phi^2}{2}}} d X_2 ,\label{NN}
\end{equation}
 where  $t_*$ corresponds to the time when the cosmological scale exits the horizon, and the time $t_f$ denotes the end of the inflationary epoch. Here we have considered that $N(t=t_f)=0$.

In this way, from Eq.(\ref{NN}) we obtain that the number of $e-$folds $N_*$ can be written as
\begin{equation}
N_*=\ln\left(\frac{\phi_*}{\phi_f}\right)-\left(\frac{1}{2}\frac{\tilde{V}_0}{\bar{V}_0}\right)\,(\mbox{Ei}[b_1X_{2_*}]-\mbox{Ei}[b_1X_{2_f}]),\label{N1}
\end{equation}
where the constant $b_1=\kappa^2[2/3-(2X_2^{(0)})^{-1}]$ and the function Ei[z] corresponds to the exponential integral, see Ref.\cite{book1}.

In order to find an analytical expression for the scalar field $\phi$ (or $X_2$) when the cosmological scale exists the horizon, we can assume small values of $X_2$ during the inflationary scenario.  In this way,  considering the relation between $\phi$ and $X_2$ given by  Eq.\eqref{relation_sum_f}, we obtain that the number of $e-$ folds from Eq.(\ref{N1}) is simplified to 
\begin{equation}
     N_* \simeq  \ln{\left(\dfrac{\phi_*^{\frac{\tilde{V}_0}{\bar{V}_0} + 1}}{\phi_f^{\frac{\tilde{V}_0}{\bar{V}_0} + 1}}\right)} ,\label{Na}
\end{equation}
and then the scalar field $\phi$ when the cosmological scale exits the horizon yields
\begin{equation}
    \phi_* \simeq \phi_f e^{\dfrac{N_*}{\frac{\tilde{V}_0}{\bar{V}_0} + 1}}.\label{FN}
\end{equation}
Also, in order to find the scalar field at the end of inflation, we can consider that  
inflation ends when the slow roll parameter $\epsilon=1$ (or equivalently $\ddot{a}=0$). Thus, from Eq.\eqref{slow_roll_epsilon}  we obtain that  the scalar field at the end of the inflationary epoch $\phi_f$ results
\begin{equation}
    \phi_f \simeq \frac{\sqrt{2} (\bar{V}_0+\tilde{V}_0)}{\kappa  \sqrt{\bar{V}_0} \sqrt{\bar{V}_0-4 \tilde{V}_0 X_2^{(0)}}}, 
    \label{FF}
\end{equation}
where the amplitude $\bar{V}_0>4\tilde{V}_0\,X_2^{(0)}$, in order to obtain a real value for the scalar field at the end of inflation. In particular,  if the potential $\bar{V}_0\gg \tilde{V}_0\,X_2^{(0}$ then the scalar field at the end of inflation is simplified  to $\phi_f\simeq [1+\tilde{V}_0/\bar{V}_0](\sqrt{2}/\kappa)$.

{To find the dependence of the fields as a function of cosmological time i.e., $\phi(t)$ and $X_2(t)$, we will first determine the relationship between the fields and the scale factor using the slow roll approximation. In this form, as we have used that $N(t)=\int_{t}^{t_f} Hdt$, with $N(t=t_f)=N_f=0$, then $a/a_f=e^{-N}$ or equivalently $N=\ln(a_f/a)$, where $a_f$  as before corresponds to the final value of the scale factor. Thus, from Eq.(\ref{Na}) we have that the relation between the scale factor and the scalar field can be written as
\begin{equation}
\phi(a)\simeq\phi_f\,\left(\frac{a}{a_f}\right)^{-1/[(\tilde{V}_0/\bar{V}_0)+1]} ,\label{fa}
\end{equation}
and considering  Eq.(\ref{relation_sum_f}) we find the relation between $X_2=X_2(a)$ and the scale factor is given by
\begin{equation}
X_2(a)=X_2^{(0)}\phi_f^2 \,\left(\frac{a}{a_f}\right)^{-2/[(\tilde{V}_0/\bar{V}_0)+1]}.
\end{equation}
Now, differentiating with respect to time Eq.(\ref{fa}), we obtain that the Hubble parameter satisfies the relation $H=-(\tilde{V_0}/\bar{V_0}+1)(\dot{\phi
}/\phi)$, in which $\dot{\phi}<0$, in order to ensure that the Hubble parameter $H>0$. Thus, replacing this expression for the Hubble parameter into Eq.(\ref{srbeq2}) and considering the effective potential given by Eq.(\ref{VS}), we find that the differential equation that relates the scalar field $\phi$ and the cosmological time becomes
\begin{equation}
\dot{\phi}=\pm\sqrt{A_0}\,\phi\,e^{(\kappa\phi/2)^2},\,\,\,\,\,\mbox{where}\,\,\,\,\,\,\,\,\,\,\,\,\,A_0=\frac{\kappa^2\,\bar{V}_0}{3(\tilde{V_0}/\bar{V_0}+1)}.\label{ft}
\end{equation}
Here, we should consider the negative sign of Eq.(\ref{ft}) when the scalar field 
$\phi$ takes positive values, and vice versa.
In this form,  from Eq.(\ref{ft}) we find that the solution of the scalar field in terms of the cosmological time can be written as
\begin{equation}
\frac{1}{2}\,\mbox{Ei}[-(\kappa\phi/2)^2]=\pm\sqrt{A_0}t+C_0,
\end{equation}
and thus,  a numerical analysis is required to determine $\phi(t)$. Here 
 the function Ei[z] corresponds to the exponential integral function  \cite{book1} and the quantity  $C_0$ denotes an integration constant. }

On the other hand, in relation to the cosmological perturbations, we find that the 
scalar power spectrum when the cosmological scale exits the horizon for these exponential potentials becomes
\begin{equation}
    \mathcal{P}_{\mathcal{R}} |_* =\mathcal{P}_{\mathcal{R}_*} \simeq \frac{16 \kappa ^2 (\bar{V}_0+\tilde{V}_0)^3}{3 \bar{V}_0 \phi_*^2 (\bar{V}_0-4 \tilde{V}_0 X_2^{(0)})} =\dfrac{8}{3} \kappa ^4 (\bar{V}_0+\tilde{V}_0) e^{-\frac{2 N_*}{\frac{\tilde{V}_0}{\bar{V}_0}+1}} .
\label{P}
\end{equation}
Here we have utilized Eqs.(\ref{PR}), (\ref{relation_sum_f}), (\ref{FN}) and (\ref{FF}), respectively.

Additionally, from Eq.(\ref{nR}) the scalar spectral index can be written in terms of the number of $e-$ folds $N$ as
$$
    n_{\mathcal{R}} |_* \simeq -\frac{\kappa ^2 \phi_*^2 \left(27 \bar{V}_0^2-124 \bar{V}_0 \tilde{V}_0 X_2^{(0)}-16 \tilde{V}_0^2 X_2^{(0)}\right)}{9 (\bar{V}_0+\tilde{V}_0)^2} 
 $$   
 \begin{equation}   
    = -\frac{2 e^{\frac{2 N_*}{\frac{\tilde{V}_0}{\bar{V}_0}+1}} \left(27 \bar{V}_0^2-124 \bar{V}_0 \tilde{V}_0 X_2^{(0)}-16 \tilde{V}_0^2 X_2^{(0)}\right)}{9 \bar{V}_0 (\bar{V}_0-3 \tilde{V}_0)}.
\label{n}
\end{equation}
Here, we have used that 
$$
\cos\theta=\frac{\sqrt{4\epsilon_\phi}}{\sqrt{9\epsilon_ A+4\epsilon_\phi}},\,\,\,\,\mbox{and}\,\,\,\,\sin\theta=\frac{\sqrt{9\epsilon_A}}{\sqrt{9\epsilon_ A+4\epsilon_\phi}},
$$
 to determine the slow roll parameter $\eta_{\sigma\sigma}$ associated to the scalar spectral index $n_{\mathcal{R}}$, see Eq.(\ref{nR}).

Besides, the tensor-scalar ratio at Hubble-crossing as a function of the number of $e-$folds $N$ results
\begin{equation}
   r |_*=r_*\simeq \frac{8 \kappa ^2 \bar{V}_0 \phi_*^2 (\bar{V}_0-4 \tilde{V}_0 X_2^{(0)})}{(\bar{V}_0+\tilde{V}_0)^2} = 16 e^{\frac{2 N_*}{\frac{\tilde{V}_0}{\bar{V}_0}+1}}\,,
\label{r}
\end{equation}
where we have used Eq.(\ref{rr}).

\begin{figure}[h!]
   \centering
       \includegraphics[scale=0.6]{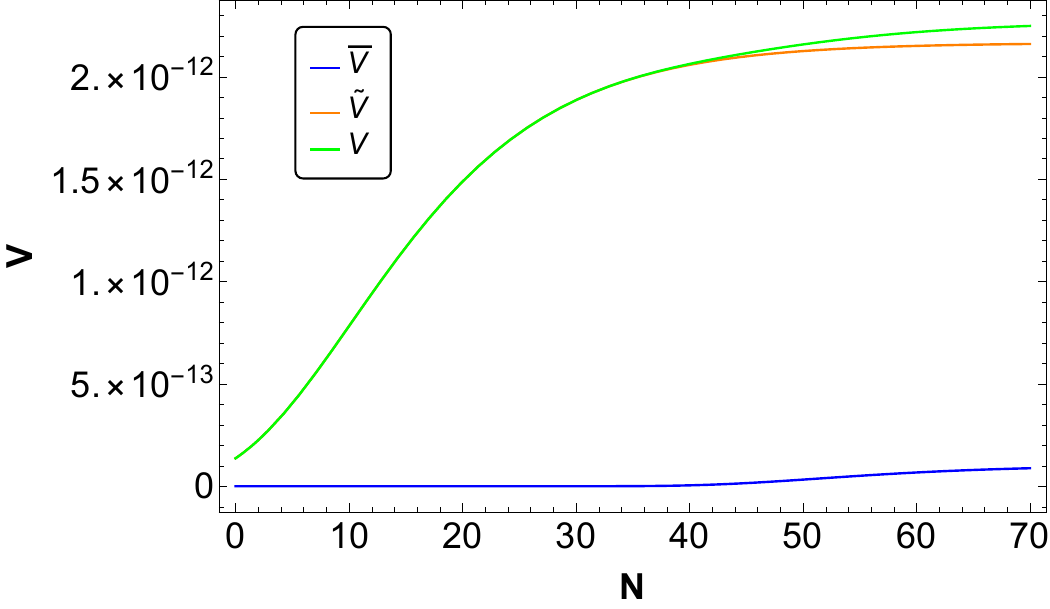}
       \caption{Evolution of the potentials $\overline{V}$ and $\tilde{V}$ and the total potential $V=\overline{V}+\tilde{V}$ in terms of the number of $e-$ folds $N$ during the inflationary epoch. Here, we have utilized the constraints on the parameters given by Eq.(\ref{Val}). Here we have considered $\kappa=1$\label{pot1} }
\end{figure}

In order to find the parameters $\bar{V}_0$, $\tilde{V}_0$ and $X_2^{(0)}$, we can consider that  the observational parameters take the following  values; the power spectrum  
$\mathcal{P}_{\mathcal{R}_*}=2.2*10^{-9}$, $n_{\mathcal{R}_*}=0.967$ and the tensor scalar ratio $r_*=0.04$ at the number of $e-$ folds $N_*=60$. In this form, we obtain the following constraints on the parameter space:
\begin{equation}
\bar{V}_0\simeq 1.03\times 10^{-13}\kappa^{-4},\,\,\,\,\,\tilde{V}_0\simeq 2.17\times 10^{-12}\kappa^{-4},\,\,\,\mbox{and}\,\,\,\,\,X_2^{(0)}\simeq-1.21\times10^{-2},\label{Val}
\end{equation}
respectively. Here we have utilized Eqs.(\ref{P}), (\ref{n}) and (\ref{r}). 

In Fig.(\ref{pot1}) we show the evolution of the potentials $\bar{V}(\phi)$, $\tilde{V}(X_2)$ and the total potential $V(\phi,X_2)$ as a function of the number of $e-$folds $N$. Here, we have considered the constraints on the space-parameter given by Eq.(\ref{Val}), which were obtained using the observational Planck data\cite{Planck2018}. In order to write down values of the  effective potentials $\bar{V}(\phi)$,  $\tilde{V}(X_2)$ and the total potential $V(\phi,X_2)$ as a function of the number of $e-$folds $N$, we have used Eqs.(\ref{VS}), (\ref{relation_sum_f}) and (\ref{FN}). 
Thus, from this plot, we can note that  $\tilde{V}(X_2)>\bar{V}(\phi)$ due to the constraints 
found on the parameter-space given by Eq.(\ref{Val}). In this way, we can conclude that the inflationary expansion of the universe is driven by the field $X_2$ from its potential $V(X_2)$, since during inflation $3H^2\simeq \kappa^2\,V\sim\kappa^2\tilde{V}(X_2)$.

{Additionally, we can note from Fig. (\ref{pot1}) that the inflationary energy scale associated with the effective potential $V\sim \tilde{V}\sim H^2\kappa^{-2}\lesssim 10^{-12}\kappa^{-4}$ during inflation,  where $N\lesssim 60$. In this respect, this energy scale  is similar to that found  in single-field chaotic inflationary models in which the potential is defined as $V\propto\phi^{\alpha}$. In particular for this potential was obtained that  in the specific case when  $\alpha=2$, the inflationary energy scale corresponds to $V\sim 10^{-14}\kappa^{-4}$ and for $\alpha=4$ the energy scale during inflation is $V\sim 10^{-12}\kappa^{-4}$
\cite{Linde:1983gd,Liddle:1993ch}. For the case of the Starobinky model, see Ref.\cite{Starobinsky:1980te}, in which the conformal potential $V=(3/4)(m^2/\kappa^2)(1-e^{-\sqrt{2/3}\phi\kappa})^2$ the observable CMB amplitude fixed the parameter $m\simeq 10^{-5}/\kappa$, with which the energy scale is $V\sim10^{-10}\kappa^{-4}$ during inflation in this model\cite{Ivanov:2021chn}. Thus, comparing with these well-known  inflationary models (chaotic and Starobinky), we can note that our model has a similar inflationary energy scale to these single-field inflation models. In relation to the end of the inflationary stage, we have considered that inflation ends when $\epsilon=1$ and then the value of the scalar field at the end inflation is given by Eq.(\ref{FF}) or equivalently  at $X_{2_f}=2X_2^{(0)}(\tilde{V}_0+\bar{V}_0)^2/(\kappa^2\bar{V}_0[\bar{V}_0-4\tilde{V}_0X_2^{(0)}])$. Here we have used the relation between both fields given by Eq.(\ref{relation_sum_f}). In this sense, from the relation defined by Eq.(\ref{relation_sum_f}) between both fields, we have the phase-space becomes one-dimensional and the end of inflation occurs   only along a single classical trajectory in the field space of $\phi$ (or $X_2$). }



\section{Example II: Effective Potential $V(\phi,X_2) =\bar{V}(\phi)\,\times \tilde{V}(X_2) $}

In this section, we will analyze an effective potential involving an interaction between the fields $\phi$ and $X_2$. This interaction is represented by the product of their individual potentials, $\bar{V}(\phi)$ and $\tilde{V}(X_2)$. In this context, we consider, as a second example,  the effective potential $V(\phi,X_2)$ given by \cite{Garcia-Bellido:1995hsq}
\begin{equation}
   V(\phi,X_2) =V= \bar{V}(\phi)\,\times\,\tilde{V}(X_2) .
  \end{equation} 
 From this effective potential, we obtain that the relation between both fields is described by the following differential equation   
\begin{equation}
    \dfrac{d \phi}{d A} = - \dfrac{2}{3 A} \dfrac{\bar{V}_{,\phi}}{\kappa^2\bar{V}} \left(\dfrac{1}{\frac{\tilde{V}_{,X_2}}{\kappa^2 \tilde{V}}-\frac{2}{3}} \right).\label{ED2}
\end{equation}

Additionally, the number of $e-$folds $N$ for this product of potentials becomes
\begin{equation}
    d N = - \kappa^2 \dfrac{\bar{V}}{\bar{V}_{,\phi}} d \phi + \dfrac{\kappa^2}{3 X_2} \dfrac{\tilde{V}}{(\tilde{V}_{,X_2}-\frac{2}{3}\kappa^2 \tilde{V} ) } d X_2 . 
\end{equation}


To apply our results and derive analytical solutions to the equations, we make the assumption that the individual potentials associated with the fields $\phi$ and $X_2$ are defined as follows \cite{Byrnes:2008wi}
\begin{equation}
\bar{V}(\phi) = \bar{V}_0\phi^{\overline{a}}, \,\,\,\,\,\,\mbox{and}\,\,\,\,\,\,\,\,\tilde{V} (X_2)= \tilde{V}_0 e^{\tilde{b} X_2}.\label{50}
\end{equation}
Here, as before, $\bar{V}_0$ and $\tilde{V}_0$ are two arbitrary constants. Additionally, the exponents $\overline{a}$ and $\tilde{b}$ associated with the power-law potential $V(\phi)$ and the exponential potential $V(X_2)$ are  constants. This type the potential  defined as $V(\phi,X_2)\propto\phi^{\overline{a}}\,e^{\tilde{b}X_2} $  is known in the literature as product-exponential (PE) potential. It was first introduced in Ref.\cite{Byrnes:2008wi} (with $\overline{a}=2$) as  quadratic times exponential potential, see also Refs.\cite{Elliston:2011dr,Huston:2013kgl,Gonzalez:2018jax}. Besides,  the product $\bar{V}_0\,\tilde{V}_0=V_0$ sets the energy scale of the PE potential $V(\phi,X_2)$.  

From the chaotic times exponential potential $V$, we observe that the differential equation described by Eq.(\ref{ED2}) yields
\begin{equation}
    \dfrac{d \phi}{d A} = \dfrac{1}{\phi A} \dfrac{\overline{a}}{\kappa^2 (1-b)} \label{relation_plus1_f},
\end{equation}
and then the relation between $\phi$ and $X_2$ is given by 
\begin{equation}
    \phi^2 = \text{ln}\left( \dfrac{X_2}{X_2^{(0)}} \right)^{n/2} ,\label{52}
\end{equation}
in which the quantity $n$ is defined as $n=\dfrac{\overline{a}}{\kappa^2 (1-b)}$ with $b\neq 1$ and $X_2^{(0)}$ corresponds to an integration constant. Here, we have considered for convenience  that  $\tilde{b}=(2/3)b\kappa^2$, in which $b$ corresponds to a dimensionless constant. 

Also, we find that the number of $e-$folds  for the PE potential can be written as
\begin{equation}
 N_*=   \int_{t_*}^{t_f} Hdt \simeq \int_{\phi_f}^{\phi_*} \dfrac{\kappa^2 \phi}{\overline{a}} d \phi + \dfrac{n \kappa^2}{2\overline{a}} \int_{X_{2f}}^{X_{2*}} \dfrac{1}{X_2}  d X_2,
\end{equation}
and using equation \eqref{52}, we can obtain  the relation between the number of $e-$folds and the scalar field $\phi$ at the end of inflation, as well as when the cosmological scale exits the horizon results 
\begin{equation}
    \phi^2_*-\phi^2_f = \frac{\overline{a}}{\kappa^2} N_*.\label{54}
\end{equation}

To determine the value of the scalar field $\phi$ at the end of inflationary stage $\phi_f$, we can assume that the slow roll parameter $\epsilon$ given by Eq.(\ref{ee}) is $\epsilon(\phi=\phi_f)=1$ with which

\begin{equation}
    \phi_f = \frac{\overline{a}}{\sqrt{2} \sqrt{\kappa ^2-2 (b-1) b \kappa ^4 X_2^{(0)}}}.\label{ff2}
\end{equation}


\begin{figure}[b!]
   \centering
       \includegraphics[scale=0.6]{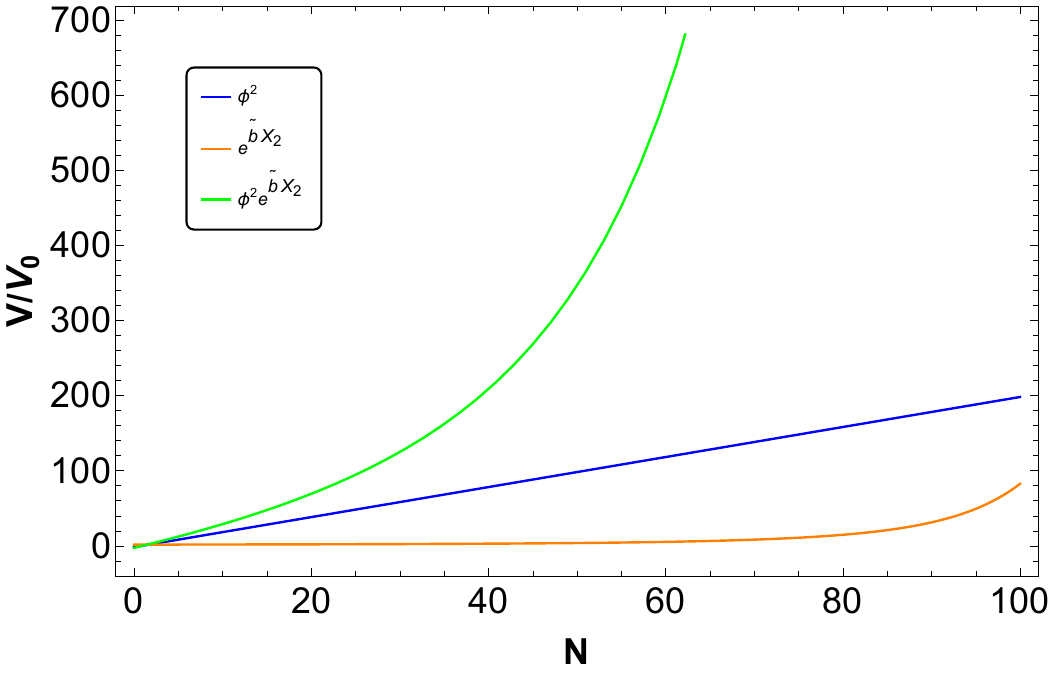}
       \caption{Evolution of the  PE  potential  $V(\phi,X_2)/V_0=\phi^2\,e^{\tilde{b}X_2}$, the terms $\phi^2$ and $e^{\tilde{b}X_2}$ as a function of the number of $e-$ folds $N$, respectively. Here we have used the values of the parameters $b$ and $X_2^{(0)}$ given by Eq.(\ref{co2}) and $\kappa=1$.}\label{F2}
\end{figure}

{As before, to find the dependence of the fields as a function of cosmological time i.e., $\phi(t)$ and $X_2(t)$, we will first determine the relationship between the fields and the scale factor using the slow roll approximation. Thus, the scalar field can be written as
\begin{equation}
\phi(a)\simeq \sqrt{\phi_f^2 - \dfrac{\overline{a}}{\kappa^2} \ln \dfrac{a}{a_f}},\label{fa2}
\end{equation}
and, by applying the last equation in Eq. (\ref{52}), we can determine \( X_2 = X_2(a) \).\\
Now, differentiating with respect to time Eq.(\ref{fa2}), we obtain that the Hubble parameter satisfies the relation $H=-(2 \kappa^2 \phi \dot{\phi}/\overline{a})$, in which $\dot{\phi}<0$, in order to ensure that the Hubble parameter $H>0$. Thus, replacing this expression for the Hubble parameter into Eq.(\ref{srbeq2}) and considering the effective potential given by Eq.(\ref{VS}), we find that the differential equation that relates the scalar field $\phi$ and the cosmological time becomes
\begin{equation}
\dot{\phi}=\pm\sqrt{A_0}\,\phi^{\frac{\overline{a}}{2}-1}\,e^{\frac{\tilde{b}}{2} X_2},\,\,\,\,\,\mbox{where}\,\,\,\,\,\,\,\,\,\,\,\,\,A_0=\frac{V_0 \overline{a}^2}{6 \kappa^2},\label{ft2}
\end{equation}
but, in order to solve this differential equation for $\phi(t)$,  we note from Eq.(\ref{52}) that it is necessary to perform a numerical analysis.
 }

On the other hand, for the PE potential, the scalar power spectrum as a function of the number of $e-$ folds $N$ from Eq.(\ref{PR}) can be written as
\begin{equation}
    \mathcal{P}_{\mathcal{R}}  \simeq  \frac{16 \kappa ^6 \bar{V}_0 \tilde{V}_0 e^{\frac{2}{3} b \kappa ^2 X_2^{(0)}} \left(\frac{\overline{a}^2}{2 \left(\kappa ^2-2 (b-1) b \kappa ^4 X_2^{(0)}\right)}+\overline{a} N\right)^{\frac{\overline{a}+2}{2}}}{3 \overline{a} \left(4 (b-1) b \kappa ^4 N X_2^{(0)}-\frac{\overline{a}}{2 (b-1) b \kappa ^2 X_2^{(0)}-1}\right)},\label{P2}
\end{equation}
and the scalar spectral index in terms of the number $N$ considering Eq.(\ref{nR}) becomes
\begin{equation}
    n_{\mathcal{R}}  \simeq \frac{4}{9} (27-23 b) b X_2^{(0)}-\frac{3\overline{a}}{\frac{\overline{a}}{2-4 (b-1) b X_2^{(0)}}+N}.\label{n2}
\end{equation}

Additionally, we find that the  tensor to scalar ratio $r$ in terms of the number of $e-$ folds $N$   can be written as 
\begin{equation}
   r \simeq \frac{8 \overline{a}^2}{\kappa ^2 \left(\frac{\overline{a}^2}{2 \left(\kappa ^2-2 (b-1) b \kappa ^4 X_2^{(0)}\right)}+\overline{a} N\right)}+32 (b-1) b \kappa ^2 X_2^{(0)}.\label{r2}
\end{equation}
Here we have used Eq.(\ref{rr}).

In the following, by simplicity, we will consider the special case in which the power $\overline{a}$ associated to the power law potential $\bar{V}(\phi)$ is equal to $\overline{a}=2$, i.e., the chaotic potential $\bar{V}(\phi)\propto \phi^2$.
Besides, in order to find the parameters $b$, $V_{0}$ ($V_{0}=\bar{V}_0 \tilde{V}_0$) and $X_2^{(0)}$ we can consider that the observational parameters; $\mathcal{P}_{\mathcal{R}}$, $n_{\mathcal{R}}$ and the tensor to scalar ratio $r$, when the number of the $e-$folds $N_*=60$.   
In this form, considering that the observational parameters take the   values from Planck data\cite{Planck2018};   
$\mathcal{P}_{\mathcal{R}_*}=2.2*10^{-9}$ for the power spectrum, $n_{\mathcal{R}_*}=0.967$ and for the tensor scalar ratio $r_*=0.04$ (upper bound) at the number of $e-$folds $N_*=60$, then  we obtain the following constraints on the parameter-space:
\begin{equation}
b\simeq 0.987,\,\,\,\,\,V_0\simeq 5.308\times 10^{-15}\kappa^{-2},\,\,\,\mbox{and}\,\,\,\,\,X_2^{(0)}\simeq 0.562,\label{co2}
\end{equation}
where we have utilized Eqs.(\ref{P2}), (\ref{n2}) and (\ref{r2}), respectively

In Fig.(\ref{F2}) we show the PE potential $V(\phi,X_2)/V_0$ together with the terms $\phi^2$ and $e^{\tilde{b}X_2}$ as a function of the number of $e-$ folds $N$. In this plot, we have used the values of $b$ and $X_2^{(0)}$ obtained assuming the observational parameters from  Planck data, see Eq.(\ref{co2}). 
In order to write down values of the PE potential $V(\phi,X_2)/V_0$ together the quantities $\phi^2$ and $e^{\tilde{b}X_2}$ as a function of the number of $e-$folds $N$, we have utilized  Eqs.(\ref{50}),(\ref{52})  and (\ref{54}) together with the constraints obtained on $b$ and $X_2^{(0)}$.  From Fig. (\ref{F2}) we note that during the inflationary epoch in which the number of $e-$ folds $N<70$, the term $\phi^2\gg e^{\tilde{b}X_2}$ and this suggests that inflation is driven by the term $\phi^2$. In this form,  we find that the dominant term on the PE potential corresponds to the quadratic term associated to the scalar field $\phi$.  In this sense, we can approximate the PE potential during inflation in which $N<70$ as $V(\phi,X_2)/V_0\sim \phi^2(1+\tilde{b}X_2+..)=\phi^2(1+\tilde{b}A^2/2+..)$ being the dominant term $\phi^2$ in the potential $V(\phi,X_2)$. It is important to note that in the literature, this approximate potential becomes fundamental in the study of the reheating scenario after inflation, see e.g. Ref.\cite{Kofman:1994rk}. 

{In relation to the inflationary energy scale in this model, we note from Fig.(\ref{F2}) that the potential $V=V(N)$ at $N\sim 60$ becomes $V\sim \mathcal{O}(10^3)V_0\kappa^{-2}$. Considering the constraint from the observational parameters in which $V_0\sim10^{-14}\kappa^{-2}$, we find that the scale the energy in this PE model becomes $ V\sim\mathcal{O}(10^{-11})\kappa^{-4}$, which  is very similar to that found in the previously analyzed model. Additionally, from Eq.(\ref{52}) we note that, as before,  the phase-space becomes one-dimensional and  inflation ends along a single classical trajectory in the field space. In this context, the value of the scalar field at the end of inflation $\phi_f$ is defined by Eq.(\ref{ff2}).}

\section{Conclusions}\label{Concluding_Remarks}
 
 In this paper we have studied an inflationary scenario during the early universe, in the context of a theory described by a scalar field together with a vector field minimal coupled to gravity. In this theory, we have obtained the background equations considering a flat FRW metric together with an effective potential associated to both fields. Besides,  we have assumed the slow roll approximation in these equations in order to find analytical solutions to the background variables. By using the local field rotation for both fields to recognize the adiabatic and entropy perturbations, we have found the new slow-roll parameters and the observational parameters such as the power spectrum, scalar spectral index, and tensor-to-scalar ratio in our theory. 

In order to apply this scenario, we have considered two different effective potentials $V(\phi,X_2)$. As a first example, we have assumed that the potential associated to both fields does not present interaction between the fields $\phi$ and $X_2$, such that, the chosen potential becomes $V(\phi,X_2)=\bar{V}(\phi)+\tilde{V}(X_2)$. As a second example, we have considered an effective potential that presents an interaction between both fields and it is given by $V(\phi,X_2)=\bar{V}(\phi)\,\times\,\tilde{V}(X_2)$.

For the first example in which the total potential $V(\phi,X_2)=\bar{V}(\phi)+\tilde{V}(X_2)$ we have considered that both potentials $\bar{V}(\phi)$ and $\tilde{V}(X_2)$ correspond to  exponential  potentials  defined  by Eq.(\ref{VS}). Under the slow-roll approximation, we have found the relation between the fields $\phi$ and $X_2$ given by Eq.(\ref{relation_sum_f}) in which $X_2\propto\phi^2$. Also, we have determined the number of $e-$
folds $N$ and the value of the scalar field at the end of the inflationary epoch, see Eqs.(\ref{Na}) and (\ref{FF}), respectively. 
In order to find the parameters $\bar{V}_0$, $\tilde{V}_0$ and $X_2^{(0)}$, we have assumed   the observational parameters in which; the power spectrum  
$\mathcal{P}_{\mathcal{R}_*}=2.2*10^{-9}$, $n_{\mathcal{R}_*}=0.967$ and the tensor scalar ratio $r_*=0.04$ at the number of $e-$ folds $N_*=60$. Here, we have determined that the parameter-space is given by Eq.(\ref{Val}). From these values on the parameters, we have found that the potential associated to the field $X_2$ is greater than $\bar{V}(\phi)$, i.e.,   $\tilde{V}(X_2)>V(\phi)$ as shown in Fig.(\ref{pot1}).  In this context, we have found that the inflationary scenario is driven by the field $X_2$ through its exponential potential $\tilde{V}(X_2)\sim V(\phi,X_2)\simeq H^2$.

As a second example we have analyzed an effective potential $V(\phi,X_2)$ defined   as the product $V(\phi,X_2)=\bar{V}(\phi)\times \tilde{V}(X_2)$. In particular, we have assumed a chaotic potential for the scalar field $\phi$ and an exponential potential for the field $X_2$ (see Eq.(\ref{50})) and this total potential is known in the literature as product-exponential potential or PE potential. By considering the slow-roll approximation, we have obtained the relation between the fields $\phi$ and $X_2$ given by Eq.(\ref{52}). Besides, we have found  the number of $e-$
folds $N$ and the value of the scalar field at the end of the inflationary epoch, as in the previous example.
As before, we have obtained the constraints on the parameter-space considering the observational parameters. In particular we have found the constraints on the parameters $b$, $V_0$ and the integration constant $X_2^{(0)}$, see Eq.(\ref{co2}). From Fig. (\ref{F2}) we have found that during the inflationary epoch in which the number of $e-$ folds $N<70$, the term $\phi^2\gg e^{\tilde{b}X_2}$ and this suggests that inflation is driven by the term $\phi^2$. In this sense, we have found that 
 the dominant term on the PE potential corresponds to the quadratic term associated to the scalar field $\phi$. In this context, we notice that 
  the PE potential during inflation can be approximate to $V(\phi,X_2)/V_0\sim \phi^2(1+\tilde{b_3}X_2+..)=\phi^2(1+\tilde{b_3}A^2/2+..)$ as the potential studied in theory of reheating after inflation, in which the interaction potential is proportional to $\phi^2\,A^2$.

Finally, in this article, we have not addressed the inflationary scenario in a scalar vector gravity theory for other types of 
 effective potential $V(\phi,X_2)$, see e.g.\cite{M1,M2}. {Besides, we have not considered the reheating  and the particle creation in our model of double inflation.  In the process of reheating
of the universe, the matter and radiation of the universe are produced generally through the
decay of the scalar field(s), while the temperature increases
in many orders of magnitude and then the universe connects with the radiation regime of
the standard big-bang model. Thus, in relation to   the reheating   associated to double inflation, we will consider the study of Refs.\cite{Kofman:1997yn,Martin:2021frd,Aoki:2022dzd,Guendelman:2020zvt} in order to describe this stage in our model. In the context of the particle creation,  we will study this process for our model following  the Refs.\cite{Ford:1986sy,Mottola:1984ar,Nilles:2001fg,Parra:2024usv}. }

 {
 We hope to return to these points in the near future.}


\clearpage

\section{Acknowledgments}
M. Gonzalez-Espinoza acknowledges the financial support of FONDECYT de Postdoctorado, N° 3230801.






\end{document}